\documentclass[12pt]{article}

\usepackage{times}

\begin{document}
\newcommand{\de}{\delta}\newcommand{\ga}{\gamma}
\newcommand{\ep}{\epsilon} \newcommand{\ot}{\otimes}
\newcommand{\be}{\begin{equation}} \newcommand{\ee}{\end{equation}}
\newcommand{\ba}{\begin{eqnarray}} \newcommand{\ea}{\end{eqnarray}}
\newcommand{\tmod}{{\cal T}}\newcommand{\amod}{{\cal A}}
\newcommand{\bemod}{{\cal B}}\newcommand{\cmod}{{\cal C}}
\newcommand{\dmod}{{\cal D}}\newcommand{\hmod}{{\cal H}}
\newcommand{\s}{\scriptstyle}
\newcommand{\einsop}{{\bf 1}}
\def\oR{R^*} \def\upa{\uparrow}
\def\R{\overline{R}} \def\doa{\downarrow}
\def\oL{\overline{\Lambda}}
\def\nn{\nonumber}
\def\nonum{\nonumber} 
 \def\dag{\dagger}
\def\beq{\begin{equation}}
\def\eeq{\end{equation}}
\def\bea{\begin{eqnarray}}
\def\eea{\end{eqnarray}}
\def\s{\sigma}
\def\th{\theta}
\def\d{\delta}
\def\ga{\gamma}
\def\l{\left}
\def\r{\right}
\def\a{\alpha}
\def\b{\beta}
\def\g{\gamma}
\def\La{\Lambda}
\def\la{\lambda}

\def\w{\overline{w}}
\def\u{\overline{u}}
\def\o{\overline}
\def\rr{\mathcal{R}}
\def\T{\mathcal{T}}
\def\N{\overline{N}}
\def\Q{\overline{Q}}
\def\L{\mathcal{L}}
\def\m{\overline{m}}
\def\n{\overline{n}}
\def\p{\overline{p}}
\def\l{\overline{l}}
\def\le{\langle}
\def\re{\rangle}
\def\dag{\dagger}
\newcommand{\reff}[1]{eq.~(\ref{#1})}

\title{\bf Integrability and exact solution of an electronic model with long 
range interactions}

\author{K.E. Hibberd\footnote{Email: keh@posta.unizar.es}\\
Departamento de F\'{\i}sica Te\'orica, Universidad de Zaragoza,\\
50009, Zaragoza, Spain \\
\\
J.R. Links\footnote{Email: jrl@maths.uq.edu.au}\\
Centre for Mathematical Physics, 
The University of Queensland, \\
QLD, 4072,  Australia}

\maketitle
\begin{abstract} We present an electronic model with long range interactions.
Through the
quantum inverse scattering method, integrability of the  model is 
established using a one-parameter family of  
typical irreducible representations of $gl(2|1)$.  
The eigenvalues of the conserved operators 
are derived in terms of the Bethe ansatz, 
from which the energy eigenvalues of the Hamiltonian are
obtained. \end{abstract}

\section{Introduction} 
The Quantum Inverse Scattering Method (QISM)  
\cite{aba} is one of the most powerful
tools in the exact study of quantum systems. 
It can be applied in a number of contexts, including both
one-dimensional systems with nearest neighbour interactions such as the
Heisenberg \cite{aba} and Hubbard \cite{shastry} models, 
and also for the analysis of
models with long range interactions such as the Gaudin Hamiltonians 
\cite{gaudin}
and extensions \cite{hkw}.
These latter constructions in particular have received renewed attention
as it has been realised that the reduced BCS model, which was recently 
proposed to describe superconducting correlations 
in metallic grains of nanoscale dimensions \cite{vondelft}, 
can be shown to
be integrable through the use of Gaudin Hamiltonians in 
non-uniform external
fields \cite{camb}. Formulating the reduced BCS model in the framework of
the QISM reproduces the exact solution originally obtained by Richardson
and Sherman \cite{rich}, 
and opens the way for the calculation of form factors
and correlation functions \cite{zlmg}. 

Motivated by this result, one can investigate to what extent the
construction can be generalised to yield new classes of models, with
some examples already given in \cite{us}. Here, we will consider a case where
an underlying {\it superalgebraic} structure (i.e., one with both bosonic and
fermionic degrees of freedom) is employed to yield an electronic model.
The supersymmetric formulation of integrable systems can be traced back
to the work of Kulish \cite{kulish}, and recently supersymmetric Gaudin
Hamiltonians have been analysed in detail in \cite{km}.  

In this article we present a Hamiltonian derived through
the QISM from a solution of the Yang-Baxter equation (YBE) associated with
a typical irreducible representation of the Lie superalgebra $gl(2|1)$, 
which has the explicit form
\beq H= \sum_j^D\ep_jn_j -g\sum_{j,k}^D \sum_{\sigma=\pm}Q^\dag_{j\s}Q_{k\s }. 
\nn \label{ham} \eeq 
Above, the energy levels $\ep_j$ are two-fold degenerate, 
$g$ is an arbitrary coupling parameter and $D$ is
the total number of distinct energy levels. Also, $n_j$ is the fermion
number operator for energy level $\ep_j$ and for
the parameters $\a_j, ~j= 1...D$ we define  
$$ Q_{j\s}= c_{j\s}\sqrt{\a_j +1}
. X_j^{ n_{j,-\s}} , $$ 
 with $X_j= \sqrt{\a_j/(\a_j+1)}.$ The
$c_{\s},c_{\s}^\dag, ~\s=\pm$, are two-fold degenerate Fermi annihilation
and creation operators.

The Hamiltonian has a similar form to the reduced BCS model \cite{vondelft}.
There, Cooper pairs are scattered into vacant energy levels
while the one particle states are blocked from scattering. In the
Hamiltonian above, there is correlated scattering depending on the
occupation numbers. One of the features of this model is that the
scattering couplings can be varied through the choice of the parameters
$\alpha_i$. 
Via the algebraic Bethe ansatz method 
and using the minimal
typical representation
of $gl(2|1)$, from which these free parameters arise, we establish the
exact solvability of the model.  Here we outline the
necessary definitions and constructions, while full details 
will be presented  elsewhere.

The Lie superalgebra $gl(2|1)$ has generators $E^i_j$, $i,j=1,2,3$ with
supercommutator relations 
$$[E^i_j,\,E^k_l]=\delta^k_j E^i_l-(-1)^{([i]+[j])([k]+[l])}
\delta^i_lE^k_j. $$ 
Above, the BBF grading $[1]=[2]=0,\,[3]=1$ is chosen and the elements are
realised in terms of the Fermi operators through (cf. \cite{bracken})  
$E^1_2=S^+=c_+^\dagger c_-,\, \\E^2_1=S^-=c_-^\dagger
c_+,\,
E^1_1=-\alpha-n_+,
\,E^2_2=-\alpha-n_-,\,E^3_3=2\alpha+n,\,\\E^1_3=Q^\dagger_+,\,E^2_3=
Q^\dagger_-,\,E^3_1=Q_+$ and $E^3_2=Q_-$, and we set $S^z=(n_+-n_-)/2$. 
The Casimir invariant of the algebra, $C= \sum_{i,j=1}^3 E^i_j\otimes E^j_i
(-1)^{[j]}$, which commutes with all the elements of $gl(2|1)$, will also
be needed, and has the eigenvalue $\xi_C= -2 \a(\a+1)$ in the above
representation. Below we let $V(\alpha)$ denote the four-dimensional
model on which the representation acts, with the basis 
$\left|+-\right>,\,\left|+\right>,\,\left|-\right>,\,\left|0\right>$. 

\section{ The Yang-Baxter equation and integrability}
\label{ybesect}

To construct the model, we use the supersymmetric formulation of the 
QISM \cite{kulish}. 
We take the following solution of
the YBE which acts on $W\otimes W\otimes V(\alpha)$, where  $W$  
denotes the three-dimensional vector
module of $gl(2|1)$, 
 \beq
R_{12}(u-v)L_{13}(u)L_{23}(v)=L_{23}(v)L_{13}(u)R_{12}(u-v)
\label{ybeq} \eeq with  
\beq 
R(u) = I\otimes I +\frac{\eta}{u}
\sum_{m,n=1}^3 (-1)^{[n]}e^m_n \otimes e^n_m \nn
,\eeq 
and the L-operator is given by
\beq
L(u)= I\otimes I + \frac{\eta}{u} \sum_{m,n=1}^3 (-1)^{[n]}e^m_n \otimes
E^n_m. \nn 
\eeq
The representations taken for the operators $E^n_m $ are as stated above,  
the variable $u$ represents the rapidity $\eta$ is arbitrary 
and $I$ is the identity operator.

By the usual procedure of the QISM, we define a transfer matrix acting on the
$D$-fold tensor product space (for distinct $\a_i$)
$V(\a_1)\otimes V(\a_2) \otimes ...\otimes V(\a_D)$
via
$$t(u)=str_0\left(G_0L_{0D}(u-\ep_D)...L_{01}(u-\ep_1)\right),$$
which gives a mutually commuting family satisfying $[t(u),\,t(v)]=0$. Above,
$str_0$ denotes the supertrace taken over the auxiliary space labelled by 0
and $G$ can be any matrix which satisfies
$[R(u),\, G\otimes G]=0.$

For the BBF grading we choose
$G=\mbox{diag}(\exp(\b\eta), \exp(\b\eta), 1)$ and by employing the algebraic
Bethe ansatz method the eigenvalues of the transfer matrix are
found to be (cf. \cite{kulish}) 
\bea \Lambda(u)&=&\exp(\b\eta) \prod_i^D
\left(1-\frac{\eta\a_i}{(u-\ep_i)}\right) \prod_j^P a(v_j-u)  \nn\\ &&~~+
\exp(\b\eta) \prod_i^D \left(1-\frac{\eta\a_i}{(u-\ep_i)} \right)\prod_j^P
a(u-v_j)  \prod_k^M a(\g_k-u)\nn\\ &&~~- \prod_i^D
\left(1-\frac{2\eta\a_i}{(u-\ep_i)}\right) \prod_j^M a(\g_j-u), \label{tme}
\eea 
where
$a(u) = 1 + \eta/u$. The parameters $v_i, w_j$ satisfy the Bethe
ansatz equations $$ \prod_k^M a(\g_k-v_j)=-\prod_i^P 
\frac{a(v_i-v_j)}{a(v_j-v_i)},~~~ \prod_i^D \frac{\g_l- \ep_i - 2 \eta
\a_i}{\g_l- \ep_i - \eta \a_i}= \exp(\b\eta) \prod_j^P a(\g_l-v_j). $$

We now introduce the operators 
\beq T_j=\lim_{u\rightarrow\ep_j}
\frac{(u-\ep_j)}{\eta^2} t(u), ~~~\mbox{ which satisfy } ~~~[T_j,\,T_k]=0.  
\label{T} \eeq 
By taking the {\it
quasi-classical} expansion $ T_j=\tau_j+o(\eta),  
$
this leads to $$\tau_j=-\b \psi_j +\sum^D_{i\neq
j}\frac{\th_{ji}}{\ep_j-\ep_i} $$ where $\th= \sum_{m,n}^3 
E^n_m\otimes E^m_n (-1)^{[m]}$
and $\psi= E^3_3$. It is easily deduced that these operators satisfy
$[\tau_j,\,\tau_k]=0 $.

Writing $K= \sum_{i,j}^D (S^+_i S_j^- + S_i^- S_j^+ + 2 S_i^z S_j ^z), $
which satisfies $[K,\,\tau_j]=0,\,\forall\, j$, 
we define the Hamiltonian as follows; 
\bea H&=& \frac 1{2\b^2} \sum_j^D (1+2\b\ep_j)  
\tau_j + \frac 1{4 \b^3} \sum_{j,k}^D \tau_j\tau_k 
 +\frac{1}{2\beta}\sum_j^D C_j - \frac {K}{2\b}
 + 2\sum_j^D \ep_j (\a_j +1)  
 \nn\\ 
&=& \frac 1{2\b} \sum_j^D \sum_{k\neq j}^D
\th_{jk} - \frac{1}{2\b}\sum_j^D (1+2\b\ep_j) \psi_j 
+ \frac 1{4\b} \sum_{i,j}^D \psi_i \psi_j
 - \frac {1}\b \sum_{j}^D \a_j (\a_j+1)\nn\\ 
&& - \frac {K}{2 \b}
+ 2\sum_j^D \ep_j (\a_j +1).\nn \eea 
The term
involving $ \th_{jk} $ may be simplified using the Casimir invariant and
the commutation relations of the algebra $gl(2|1)$ 
\bea \sum_j^D
\sum_{k\neq j}^D \th_{jk} &=& \sum_{k, j}^D \th_{jk} - \sum_j^D C_j,\nn\\
&=& K -\frac 12 \sum_{j, k}^D \psi_j\psi_k +\sum_j^D \psi_j - 2
\sum_{j, k}^D\sum_{\sigma=\pm}Q^+_{j\sigma} Q_{k\sigma} +2\sum_{j}^D 
\a_j(\a_j+1).
\nn \eea 
For $g= 1/\b$
we obtain the Hamiltonian (\ref{ham}),  which establishes 
integrability since\\
$[H,\tau_j]=0,\,\,\forall\,j$.  

From (\ref{tme},\ref{T}) 
we obtain the eigenvalues of $\tau_j$ for the BBF grading,
\bea
\la_j &=& - 2 \b \a_j +\a_j \sum_i^M \frac 1{\g_i- \ep_j }
-2 \sum_{i\neq j}^D \frac {\a_j \a_i}{\ep_j-\ep_i}, \label{la}  
\eea
as the quasi-classical limit of the eigenvalues of the transfer matrix.  
The corresponding Bethe ansatz equations are
\beq
\b + \sum^P_j \frac 1{\g_l - v_j}
=\sum^D_i \frac {\a_i}{\ep_i-\g_l} ,~~~
\sum^M_l \frac 1{\g_l - v_j}
=2 \sum_{i\neq j}^P \frac 1{v_i-v_j}.\label{bae}
\eeq
For a given solution of the Bethe ansatz equations we find that 
the number of electrons, $N=2D-M$,  $n_+-n_-=M-2P$ and the eigenvalue of
$K$ reads 
\bea \xi_K=\frac{1}{2}(M-2P)(M-2P+2). \label{su2} \eea  
The energy eigenvalues can be computed using (\ref{la},\ref{bae},\ref{su2}) 
and are given by 
$$E=2\sum_j^D\ep_j-\sum_l^M \g_l-2g\sum_j^D \alpha_j-gM. $$ 
Similar results have been obtained for the FBB and BFB gradings, 
which will appear elsewhere.

\section*{Acknowledgements}

We thank Petr Kulish for useful discussions. Jon Links acknowledges the
Australian Research Council for financial support and Katrina Hibberd is
supported by project number BFM2000-1057 from the Ministerio de Ciencia y
Tecnologia, Spain.


\end{document}